\documentclass[aoas,MSNbibl,nameyear,seceqn,dvips]{arximspdf}
\usepackage{multirow}
\usepackage{graphicx}
\usepackage{breakurl}

\doi{10.1214/10-AOAS379}
\volume{5}
\issue{1}
\pubyear{2011}
\firstpage{381}
\lastpage{399}

\makeatletter
\newcommand{\eqref}[1]{(\ref{#1})}
\makeatother

\begin{document}
\begin{frontmatter}

\title{Construction and evaluation of classifiers for forensic document
analysis\thanksref{T1}}

\runtitle{Forensic handwriting classifiers}

\thankstext{T1}{Supported in part under a contract award from the
Counterterrorism and Forensic Science
Research Unit of the Federal Bureau of Investigation's Laboratory
Division. Names of commercial
manufacturers are provided for information only and inclusion does not
imply endorsement by the FBI.
Points of view in this document are those of the authors and do not necessarily
represent the official position of the FBI or the US Government.}

\begin{aug}
\author[A]{\fnms{Christopher P.} \snm{Saunders}\corref{}\thanksref{t2}\ead[label=e1]{csaunde6@gmu.edu}},
\author[B]{\fnms{Linda J.} \snm{Davis}\thanksref{t1}\ead[label=e2]{ldavisc@gmu.edu}},
\author[B]{\fnms{Andrea C.} \snm{Lamas}\thanksref{t1}\ead[label=e3]{alamas@gmu.edu}},
\author[B]{\fnms{John J.} \snm{Miller}\thanksref{t1}\ead[label=e5]{jmiller@gmu.edu}}
\and
\author[A]{\fnms{Donald T.} \snm{Gantz}\thanksref{t1}\ead[label=e4]{dgantz@gmu.edu}%
\ead[label=u1,url]{http://statistics.gmu.edu/}%
\ead[label=u2,url]{http://statistics.gmu.edu/pages/forensicslab.html}}

\thankstext{t2}{Supported by IC Post Doctorial Research Fellowship,
NGIA HM1582-06-1-2016.}

\thankstext{t1}{Supported by Gannon Technologies Group.}

\runauthor{C. P. Saunders et al.}

\affiliation{George Mason University}

\address[A]{C. P. Saunders\\
D. T. Gantz\\
Document Forensics Laboratory\\
Department of Applied Information Technology\\
George Mason University\\
4400 University Drive, MS 1G8\\
Fairfax, Virginia 22030\\USA\\
\printead{e1}\\
\phantom{E-mail:} \printead*{e4}\\
\printead{u2}}

\address[B]{L. J. Davis\\
A. C. Lamas\\
J. J. Miller\\
Department of Statistics\\
George Mason University\\
4400 University Drive, MS 4A7\\
Fairfax, Virginia 22030\\USA\\
\printead{e2}\\
\phantom{E-mail:} \printead*{e3}\\
\phantom{E-mail:} \printead*{e5}\\
\printead{u1}}

\end{aug}

\received{\smonth{5} \syear{2008}}
\revised{\smonth{6} \syear{2010}}

%
\begin{abstract}In this study we illustrate a statistical approach to
questioned document examination. Specifically, we consider the
construction of three classifiers that predict the writer of a sample
document based on categorical data. To evaluate these classifiers, we
use a data set with a large number of writers and a small number of
writing samples per writer. Since the resulting classifiers were found
to have near perfect accuracy using leave-one-out cross-validation, we
propose a novel Bayesian-based cross-validation method for evaluating
the classifiers.
\end{abstract}



\end{frontmatter}
%

\section{Introduction}\label{sec1}

A common goal of forensic handwriting examination is the determination,
by a forensic document examiner, of which individual is the actual
writer of a given document. Recently, there has been a growing interest
in the development of forensic handwriting biometric systems that can
assist with this determination process. Forensic handwriting biometric
systems tend to focus on two main tasks. The first task, known as
writer verification, is the determination of whether or not two
documents were written by a single writer. The second task, commonly
referred to as handwriting biometric identification, is the selection
from a set of known writers of a short list of potential writers for a
given document. (Another example of a biometric identification problem
in forensics is searching fingerprint databases to find a match for a
latent fingerprint.)

In this paper we focus on closed-set biometric identification, which
assumes that the writer of a document of unknown writership is one of
$W$ known writers with handwriting styles that have been modeled by the
biometric system. It is important to note that the fundamental forensic
writer identification problem, which is to verify that a document of
questioned writership came from a ``suspect'' to the exclusion of all
other possible writers, is not addressed in this paper. The ``exclusion
of all other possible writers'' requires an assumption that the suspect
writer has a unique handwriting profile and, further, that the
handwriting quantification contains enough information to uniquely
associate the writing sample of unknown writership with the suspect's
writing profile. These issues are addressed in handwriting
individuality studies. [See \citet{r10} and related discussion papers
in the \textit{Journal of Forensic Sciences}.] Ongoing research by \citet{n1} explores some of the issues associated with studying handwriting
individuality using computational biometric systems.

At a basic level, closed-set biometric identification is similar to a
traditional multi-group statistical discriminate analysis problem. In
this paper, we implement three different discriminant functions (or
classification procedures) for categorical data resulting from the
quantification of a handwritten document. We determine the accuracy of
these three classification procedures with respect to a database of 100
writers provided by the FBI. Each of the three classification
procedures is shown to identify with close to $100\%$ accuracy the
writer of a short handwritten note.

The quantification technology used in this study is a derivative of the
handwriting biometric identification system developed and implemented
by the Gannon Technologies Group and the George Mason University
Document Forensics Laboratory. Components of the system are described
as needed. For a document of unknown writership, the system returns a
short list of potential writers from a set of known writers. This
functionality is the common goal of most forensic biometric systems
[\citet{n3}]. A forensic document examiner can pursue a final
determination of whether someone on the short list is the actual writer
of the document of unknown writership. Throughout this paper we
restrict the short list to contain one potential writer.

In Section \ref{sec2} we provide a brief overview of statistical methods for
handwriting identification. In Section \ref{sec3} we describe the nature of the
categorical data that arises from the processing of a handwriting
sample. In Section \ref{sec4} we describe three proposed classifiers and their
construction. In Section \ref{sec5} we summarize a traditional leave-one-out
cross-validation (LOOCV) used to evaluate the classifiers on their
ability to correctly predict writership of an unknown document. All
three classifiers have near perfect classification rates using a LOOCV
scheme. In Section~\ref{sec6} we implement a LOOCV with a predictive
distribution to generate new pseudo-random writing samples based on the
left-out document for which writership is to be predicted. The
pseudo-simulation allows us to compare our classifiers and estimate the
accuracy of the classifiers as a function of the size of the document
of unknown writership. In Section \ref{sec7} we summarize our results from the
two cross-validation studies and discuss ongoing and future research.

\section{Review of handwriting identification}\label{sec2}
As illustrated by the case of the Howland Will in 1868, the statistical
interpretation of handwriting evidence has a long history in the
American legal system. [See \citet{howlandwill} for an overview.]
However, \citet{n3} report that handwriting analysis as practiced by
forensic experts is considered to be subjective, opening the field to
criticism. They state that the study of computationally-based methods
``is important both to provide tools to assist the evaluation of
forensic evidence but also to bring investigative possibilities based
on handwriting'' [\citet{n3}]. The recent National Research Council
report on the needs of the forensic sciences has pointed out that
computer-based studies of handwriting ``suggest that there may be a
scientific basis for handwriting comparison, at least in the absence of
intentional obfuscation or forgery'' [\citet{nas}].

The discussion of forensic handwriting identification, including
computa- tionally-based methods, has been vigorous. The paper of \citet
{r10} and related discussion papers give the interested reader insight
into this discussion. Of the problems in computationally-based
handwriting analysis, closed-set identification procedures have been
the most commonly studied. \citet{r11} and \citet{r8} both provide
comprehensive up-to-date literature reviews on this research area.

According to \citet{r11}, handwriting identification is usually
approached from the paradigm of statistical pattern recognition or
discriminant analysis. The most common approach to writer
identification is the building of a nearest-neighbor classifier based
on an appropriate metric for the features considered. [See, for
example, \citet{r10}, \citet{r12}, \citet{r14}, \citet{r20} and
\citet{r18}.] Using a nearest-neighbor classifier, a document of
unknown writership is classified as having been written by the writer
with the most similar writing sample in the database.

When studying larger data sets of writers, computational restrictions
may require application of two different classifiers together. This
approach involves building a fast, but not necessarily accurate,
identification procedure to generate a smaller subset of possible
writers for a document of unknown writership and then applying a more
computationally-intense method with a higher accuracy to reduce the
subset to a single writer (or short list). For example, \citet{r10}
use two nearest-neighbor classifiers, each corresponding to a different
quantification procedure, applied to the same documents. Their method
uses the first quantification to pick the 100 most similar writers in a
database of 975 writers and then uses the second quantification to
select the best writer from the 100.

\citet{r17} use weighted Euclidean distance classifiers applied to
bitmaps of character images for writer identification. \citet{r18} use
a $k$-nearest-neighbor classifier and compare it to a weighted
Euclidean distance classifier; the weighted Euclidean distance
classifier out-performed the $k$-nearest-neighbor classifier.

\citet{r11} and \citet{r12} segment writing samples into graphemes.
Then they apply clustering algorithms to the graphemes to define either
a feature space or the bins of a probability distribution. When a new
document is investigated, each grapheme is associated with an
identified cluster. This reduces the new document to a frequency
distribution describing the number of times that clusters are observed
in the new document. \citet{r11} use an information retrieval
framework to measure the proximity of a test document to those in the
training set by computing the normalized inner product of the feature
vectors. \citet{r12} calculate the chi-squared distance between the
probability distributions of a test document and each training document.

In a recent paper \citet{r14} fuse the grapheme-based features with
textural features, of which the directions of contours and run-lengths
of white pixels form probability distributions for use in calculating
chi-squared distances. While the grapheme-based features perform better
than the textural features alone, fusing distances measured across
different features yields the best results.

\citet{r11} provide a summary of the performance of the various
identification methods applied to different databases of handwriting
samples. The \citet{r22} method out-performs the other methods; the
correct writer of an unknown document out of 150 possible writers is
returned, on a short list of one, 95$\%$ of the time. This method has
been improved upon in the more recent research by \citeauthor{r9} (\citeyear{r9,r3}) and
applied to much larger data sets than the initial 150 writer study.

\section{Quantification, samples and processing}\label{sec3}
\subsection{Isomorphic graph types and isocodes}\label{sec3.1}
The recent research of \citet{r2} reports that representing each
character as a ``\textit{graphical isomorphism}'' provides significant
potential to identify the writer of an unknown document. The graphs are
mathematical objects consisting of edges (links) and vertices (nodes).

\begin{figure}

\includegraphics{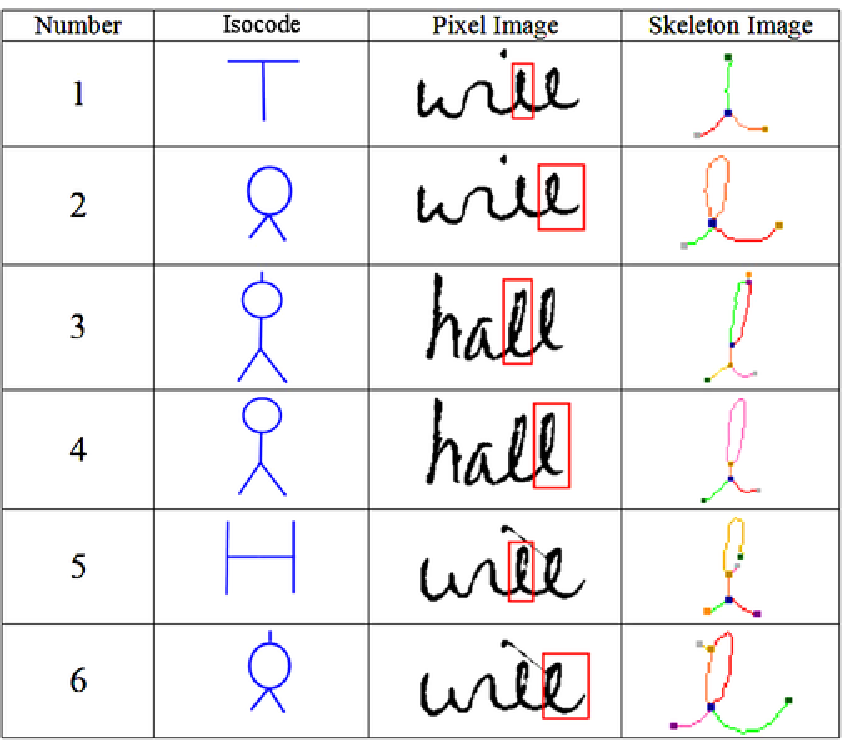}

\caption{Several isocodes used to represent the lowercase ``l.''
Comments on figure:
number 1 occurs because the writer did not make a loop with white space.
Number 2 is the copybook form for a lowercase ``L.''
Number 3 occurs because the writer filled in the loop enough at the
bottom for the skeletonizer to create a line segment at the bottom of
the loop and the writer had pen drag to leave a ``hair'' near the top
of the loop.
Number 4 occurs for the same reason as 3 but without the hair at the top.
Number 5 occurs because of pen skip which breaks the loop on the right
side. The skeleton can be ``unwound'' into the H shape.
Number 6 occurs because the pen drag to the dot on the I leaves a hair
on the loop.}\label{Figure1d}
\end{figure}

\begin{figure}

\includegraphics{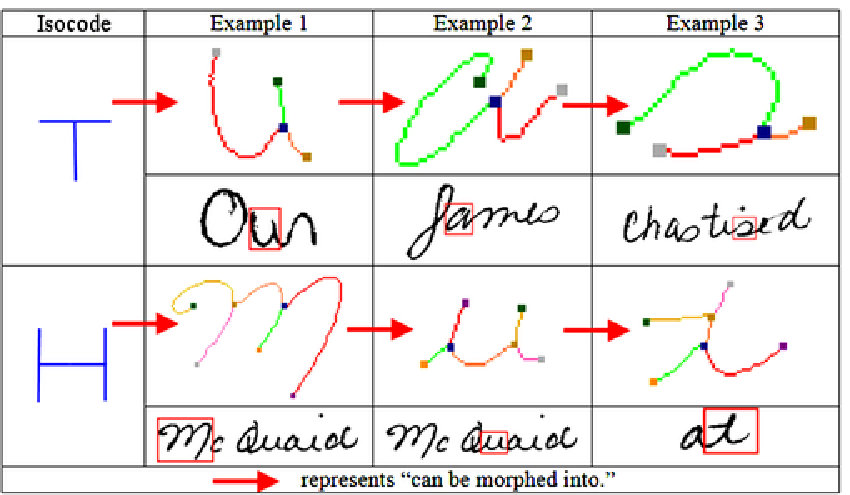}

\caption{Isomorphic graph class examples.}
\label{Figure1}
\end{figure}

The first step in the quantification of handwritten text is to convert
paper documents into electronic images. Once images are captured
electronically, individual characters are segmented either through
manual markup or automated letter recognition. (Throughout this paper,
\textit{letter} refers to the type of character and \textit{character} to
an individual instantiation of a letter. For example, ``moon'' is a word
made up of three letters and four characters.) A segmented character is
then converted to a one pixel wide skeleton. Each skeleton is then
represented by a planar graph schematic, and every schematic is
identified as belonging to a unique isomorphic class of graphs. We
refer to the isomorphic class as the \textit{isocode}. (See Figure \ref{Figure1d}.) Any two isomorphic graphs can be smoothly transformed
into one another. A particular graph, appropriately flexed and shaped,
can fit many different letters of the alphabet. Figure \ref{Figure1}
illustrates how a single isomorphic graph can represent multiple
letters by appropriate transformation.

\begin{figure}[b]

\includegraphics{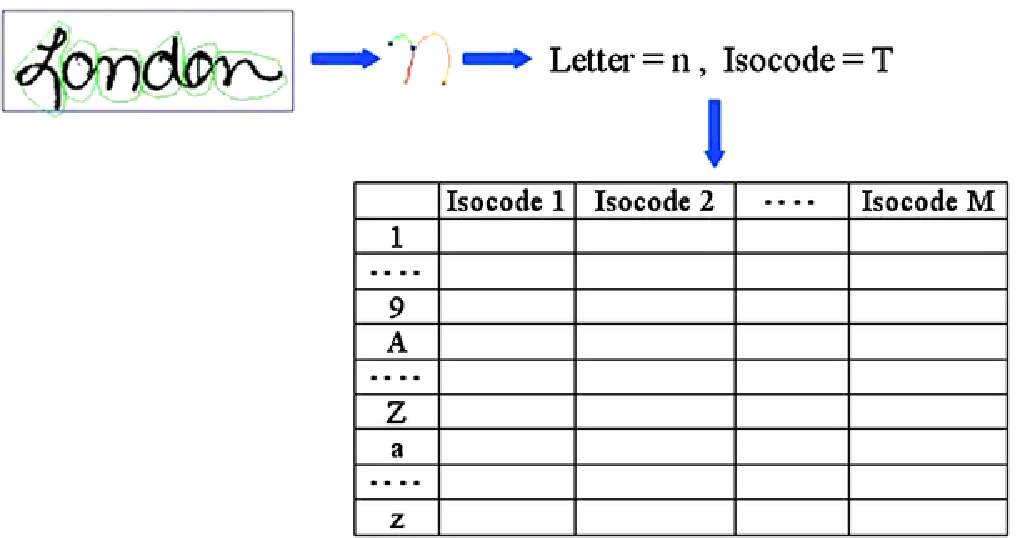}

\caption{Quantification example.}
\label{Figure1b}
\end{figure}

Recognition of a character as a particular letter and identification of
its graph as a particular isocode create an instance of a
letter/isocode pair. Each document can be represented as a matrix of
counts of the number of times each isocode is used to represent each
letter (Figure \ref{Figure1b}). The quantity of writings available
from the writer will determine the number of occurrences of any
letter/isocode pair.

The primary writer identification system described in \citet{r2} uses
an extensive set of measurements dependent on the isomorphism selected;
however, these measurements are not used in this paper. They also
report that, when the writing samples from writers are sufficiently
rich, the patterns of letter/isocode associations alone can be a
powerful identifier of writership. In our paper it is shown that the
frequencies of letter/isocode pairs provide a straightforward summary
of the data which captures sufficient information about an individual
writer to allow for accurate handwriting identification. Once the
letter/isocode pairing is done, this information can be used to
identify the most likely writer of a document (of unknown writership)
from a pool of known writers.

\subsection{Handwriting samples}\label{sec3.2}

\begin{figure}\vspace*{12pt}
\begin{quote}
Our London business is good, but Vienna and Berlin are quiet. Mr. D.
Lloyd has gone to Switzerland and I hope for good news. He will be
there for a week at 1496 Zermott St. and then goes to Turin and Rome
and will join Col. Parry and arrive at Athens, Greece, Nov. 27th or
Dec. 2nd. Letters there should be addressed 3580 King James Blvd. We
expect Charles E. Fuller Tuesday. Dr. L. McQuaid and Robert Unger,
Esq., left on the ``Y.X. Express'' tonight. My daughter chastised me
because I didn't choose a reception hall within walking distance from
the church. I quelled my daughter's concerns and explained to her that
it was just a five minute cab ride \& it would only cost $\$6.84$ for
this zone.
\end{quote}
\caption{The modified ``London Letter.''}
\label{Figure2}
\end{figure}

\begin{figure}[b]

\includegraphics{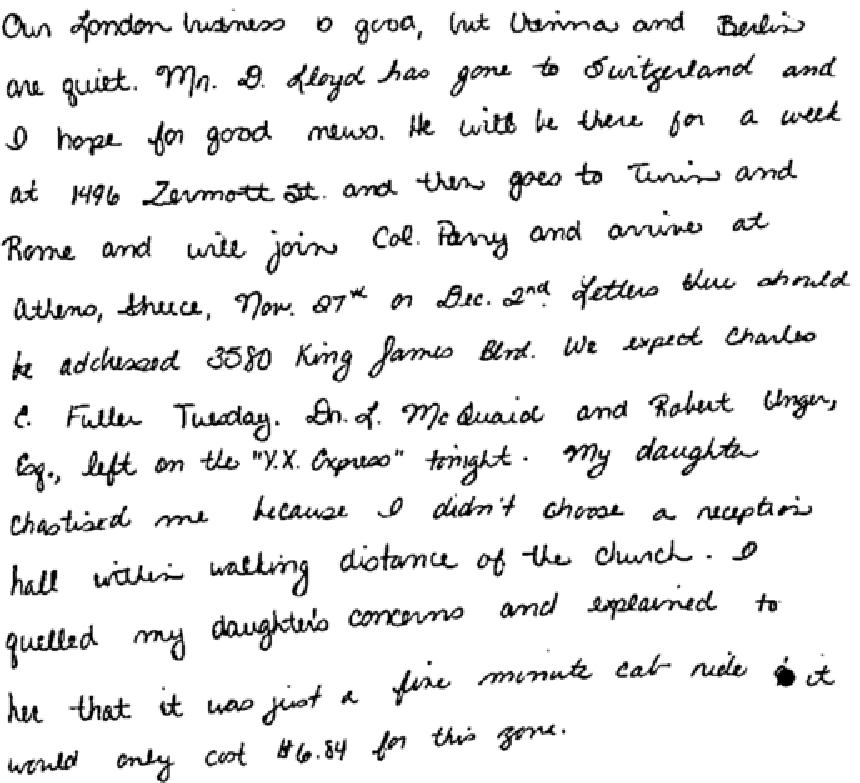}

\caption{A handwriting sample.}
\label{Figure3}
\end{figure}

The FBI conducted a project whereby writing samples were collected from
volunteers at the FBI, training classes and various forensic
conferences over a two-year period. Handwriting samples were collected
from about 500 different writers. Each writer was asked to provide 10
samples (5~in print and 5 in cursive) of a modified ``London Letter''
paragraph. (See Figures~\ref{Figure2}~and~\ref{Figure3}.)

\begin{table}
\caption{Frequency of occurrence of letters/numbers in the modified
``London Letter''}\label{parset1}
\begin{tabular*}{280pt}{@{\extracolsep{\fill}}lc@{\qquad}cc@{}}
\hline
\multicolumn{1}{@{}l}{\textbf{Letter}} & \multicolumn{1}{c}{\textbf{Frequency}}\qquad &
\multicolumn{1}{c}{\textbf{Letter}} & \multicolumn{1}{c@{}}{\textbf{Frequency}} \\
\hline
$A$ & 1 & $a$ & 35 \\
$B$ & 2 & $b$ & \phantom{0}7 \\
$C$ & 2 & $c$ & 15 \\
$D$ & 3 & $d$ & 31 \\
$E$ & 3 & $e$ & 65 \\
$F$ & 1 & $f$ & \phantom{0}6 \\
$G$ & 1 & $g$ & 10 \\
$H$ & 1 & $h$ & 23 \\
$I$ & 3 & $i$ & 28 \\
$J$ & 1 & $j$ & \phantom{0}2 \\
$K$ & 1 & $k$ & \phantom{0}2 \\
$L$ & 4 & $l$ & 22 \\
$M$ & 3 & $m$ & \phantom{0}7 \\
$N$ & 1 & $n$ & 37 \\
$O$ & 1 & $o$ & 35 \\
$P$ & 1 & $p$ & \phantom{0}5 \\
$Q$ & 1 & $q$ & \phantom{0}3 \\
$R$ & 2 & $r$ & 34 \\
$S$ & 2 & $s$ & 29 \\
$T$ & 2 & $t$ & 40 \\
$U$ & 1 & $u$ & 17 \\
$V$ & 1 & $v$ & \phantom{0}4 \\
$W$ & 1 & $w$ & \phantom{0}9 \\
$X$ & 1 & $x$ & \phantom{0}3 \\
$Y$ & 1 & $y$ & \phantom{0}6 \\
$Z$ & 1 & $z$ & \phantom{0}2 \\
$1$ & 1 & $2$ & \phantom{0}2 \\
$3$ & 1 & $4$ & \phantom{0}2 \\
$5$ & 1 & $6$ & \phantom{0}2 \\
$7$ & 1 & $8$ & \phantom{0}2 \\
$9$ & 1 & $0$ & \phantom{0}1 \\
\hline
\end{tabular*}
\end{table}

The modified ``London Letter'' paragraph used in this study includes 14
instances of numbers, 42 of uppercase letters and 477 of lowercase
letters for a total of 533 characters. (Punctuation and special
characters are ignored.) The breakdown of the frequencies of each
letter/number in the modified ``London Letter'' paragraph is given in
Table \ref{parset1}. Note that the modified ``London Letter'' is a
generalization of the standard London Letter used in collecting writing
exemplars from suspect writers.

\subsection{Processing of the FBI samples}\label{sec3.3}
The segmentation of each paragraph into characters was performed
manually by the Gannon Technologies Group, as was the association of a
letter with each character. Because the text of the paragraph is known,
the association of letters to characters should be 100$\%$ accurate.
Since some writers misspelled words and some individuals committed
errors in segmentation, the association of letters to characters was
not 100$\%$ accurate. A post-analysis of the association indicated that
the error rate in character association is less than 1$\%$.

Not all of the collected samples were processed and available for use
in this study. As a part of another study that analyzed micro features,
the cursive writing samples from the first 100 writers were divided
into two separate data sets. One of these sets (hereafter referred to
as the ``FBI 100'' data set), consisting of the first three cursive
paragraphs for these 100 writers, was available for use in this study,
resulting in a total of 293 documents. The missing paragraphs are due
to some writers' failure to submit all five of the requested cursive paragraphs.

Not all characters from each writing sample were available for use in
this study. There are three reasons for this: (a) some writers did not
submit complete paragraphs; (b) issues involving missing data in the
micro feature data (not used in this study) caused some characters to
be omitted from the data presented to us; and (c) the usage of the
first three paragraphs in the micro feature based study required the
deletion of some infrequently occurring letter/isocode pairs. The
resulting reduced number of characters per document ranged from a
minimum of 16 to a maximum of 315, with the median number of characters
per document being 160. Table \ref{parset2} summarizes the number of
characters per document. This study used all 68 isocodes in the
available data.

\begin{table}
\caption{Number of characters available in each paragraph. ID refers to
writer identifier. ``A,'' ``B,'' ``C'' refer to the three
paragraphs}\label{parset2}
\begin{tabular*}{1\textwidth}{@{\extracolsep{\fill}}lccccccccccc@{}}
\hline
\multicolumn{1}{@{}l}{\textbf{ID}} & \multicolumn{1}{c}{\textbf{A}} & \multicolumn{1}{c}{\textbf{B}} &
\multicolumn{1}{c}{\textbf{C}}&
\multicolumn{1}{c}{\textbf{ID}} & \multicolumn{1}{c}{\textbf{A}} & \multicolumn
{1}{c}{\textbf{B}} & \multicolumn{1}{c}{\textbf{C}}&
\multicolumn{1}{c}{\textbf{ID}} & \multicolumn{1}{c}{\textbf{A}} & \multicolumn
{1}{c}{\textbf{B}} & \multicolumn{1}{c@{}}{\textbf{C}} \\
\hline
\phantom{0}1 & 104& 125& 124 & 34& 171& 153& 170&\phantom{0}67& 144& 122& 139 \\
\phantom{0}2 & 156& 117& 150 & 35& 185& 156& &\phantom{0}68& 140& 142& 130 \\
\phantom{0}3 & 195& 212& 209 & 36& 292& 315& 256&\phantom{0}69&\phantom{0}41&\phantom{0}37& \phantom{0}30 \\
\phantom{0}4 & 211& 264& 237 & 37& 152& & 131&\phantom{0}70&\phantom{0}23&\phantom{0}16& \phantom{0}16 \\
\phantom{0}5 & 163& 154& 150 & 38& 201& 191& 208&\phantom{0}71& 103& 146& 123 \\
\phantom{0}6 & & 122& 130 & 39& 206& 204& 205&\phantom{0}72& 114& 117& 117 \\
\phantom{0}7 & 135& & 138 & 40& 268& 259& 261&\phantom{0}73&\phantom{0}98& 111& 128 \\
\phantom{0}8 & 162& 174& 166 & 41& 144& 156& 162&\phantom{0}74& 113&\phantom{0}91& 128 \\
\phantom{0}9 & 149& 143& 195 & 42& 286& 229& 247&\phantom{0}75& 160& 143& 148 \\
10&\phantom{0}71&\phantom{0}85& 79 & 43& 191& 191& 180&\phantom{0}76& 131& 126& 141 \\
11& 154& 160& 171 & 44& 146& 152& 132&\phantom{0}77& 149& 138& 131 \\
12& 199& 224& 217 & 45& 275& 269& 275&\phantom{0}78&\phantom{0}98&\phantom{0}96& \phantom{0}84 \\
13& 169& 169& 170 & 46& 126& 117&\phantom{0}94&\phantom{0}79& 204& 231& 204 \\
14& 206& 192& 230 & 47& 236& 184& 240&\phantom{0}80& 108& 124& 125 \\
15& 157& 143& 139 & 48& 179& 165& 184&\phantom{0}81&\phantom{0}61&\phantom{0}51& \phantom{0}53 \\
16&\phantom{0}84& & & 49&\phantom{0}86& 102&\phantom{0}97&\phantom{0}82& 115&\phantom{0}93& 102 \\
17& 193& 187& 213 & 50& 231& 215& 214&\phantom{0}83& 105& 129& 131 \\
18& 178& & 153 & 51& 197& 238& 195&\phantom{0}84& 182& 181& 171 \\
19& 260& 249& 251 & 52& 173& 166& 184&\phantom{0}85&\phantom{0}57&\phantom{0}65& \phantom{0}77 \\
20& 250& 191& 260 & 53& 257& 267& 261&\phantom{0}86& 149& 139& 125 \\
21& 208& 231& 242 & 54&\phantom{0}65&\phantom{0}84&\phantom{0}96&\phantom{0}87&\phantom{0}87& 105& 104 \\
22& 228& 186& 181 & 55& 147& 165& 139&\phantom{0}88& 147& 159& 160 \\
23& 154& 176& 168 & 56& 223& 211& 186&\phantom{0}89& 172& 166& 165 \\
24& 186& 184& 179 & 57& 163& 167& 151&\phantom{0}90& 213& 191& 208 \\
25& 163& 170& 190 & 58& 203& 229& 218&\phantom{0}91& 170& 173& 206 \\
26& 242& 216& 185 & 59& 116& 137& 130&\phantom{0}92& 178& 159& 152 \\
27& 182& 210& 187 & 60& 122&\phantom{0}99& 109&\phantom{0}93& 187& 206& 174 \\
28& 101& 111& \phantom{0}98 & 61& 116& 106& 100&\phantom{0}94& 109&\phantom{0}99& 120 \\
29& 191& 198& 200 & 62& 112& 133& 116&\phantom{0}95&\phantom{0}76&\phantom{0}50& \phantom{0}49 \\
30& 211& 222& 212 & 63&\phantom{0}95&\phantom{0}86&\phantom{0}96&\phantom{0}96& 148& 153& 158 \\
31& 167& 149& 176 & 64& 124& 123& 143&\phantom{0}97& 102&\phantom{0}89& 112 \\
32& 191& 208& 193 & 65& 185& 200& 171&\phantom{0}98& 149& 144& 155 \\
33&\phantom{0}57&\phantom{0}55&\phantom{0}66 & 66& 172& 181& 170&\phantom{0}99& 150& 149& 151 \\
& & & & & & & & 100& 152& 170& 177 \\
\hline
\end{tabular*}
\end{table}

\section{Classifiers}\label{sec4}

To facilitate this discussion, denote the number of times the~$m$th
isocode is used to write the $l$th
letter in the $j$th document written by the $i$th writer as
$n_{ijml}$, where $i = 1, 2, \ldots,W$; $j = 1, 2, \ldots,J_i $; $m
= 1, 2, \ldots,M$; and $l = 1,  2, \ldots,L.$

Let ${\mathbf{n}}_{ijl} = ( {n_{ijml} } )_{M \times1}$
denote the vector of counts corresponding to the $l$th letter in
the $j$th document written by the $i$th writer. The table of
letter/isocode frequencies for the $j$th document written by the
$i$th writer is denoted as ${\mathbf{D}}_{ij} = [ {{\mathbf
{n}}_{ijl} } ]_{M \times L}$. Let ${\mathbf{C}} \in\mathbb{N}_0
^{M \times L}$ be a matrix of nonnegative integers and let ${\mathbf
{c}}_l = ( {c_{ml} } )_{M \times1} \in\mathbb{N}_0 ^M $ be
the vector corresponding to the $l$th column. We denote the
probability of observing the matrix of counts, ${\mathbf{C}}$, in a
document written by the $i$th writer as $P( {{\mathbf{C}}|w =
i} )$, where $w$ is used to denote writer. In general, a ``$\cdot
$'' in place of a subscript denotes the summation over the dotted
subscript; for example, $n_{ij \cdot l} = \sum _{m = 1}^M
{n_{ijml} }$.

For a given document of unknown writership, say, the $v$th document
from the $u$th unknown writer, denote the corresponding counts of
isocodes used to write each letter in the document as ${\mathbf
{D}}_{uv} = [ {{\mathbf{n}}_{uvl} } ]_{M \times L}$ where
${\mathbf{n}}_{uvl} = ( {n_{uvml} } )_{M \times1} $ is the
vector of counts of isocodes used to write the $l$th letter.

Let $p_{iml}$ denote the probability of observing the $m$th
isocode given the $i$th
writer is writing the $l$th letter. We assume that ${\mathbf
{n}}_{ijl} $, $i = 1, 2, \ldots,W$, $j = 1, 2, \ldots,J_i$, and $l
= 1, 2, \ldots,L$, are independent multinomial random vectors with
parameter vectors ${\mathbf{p}}_{il} = ( {p_{iml} } )_{M
\times1} $, $p_{i \cdot l} = \sum _{m = 1}^M {p_{iml} = 1} $.
Then, under an independence assumption between letters, we have that
the probability of observing a matrix of counts, ${\mathbf{C}},$
written by the $i$th known writer is
\begin{eqnarray}\label{eq4.1}
P( {{\mathbf{C}}|w = i } ) &=& \prod _{l = 1}^L
{P( {{\mathbf{c}}_l |w = i,\mbox{ } \mathit{letter} = l} )}
\nonumber
\\[-8pt]
\\[-8pt]
\nonumber
&=& \prod _{l = 1}^L {P( {{\mathbf{c}}_l |{\mathbf{p}}_{il}
} )} ,
\end{eqnarray}
where $P( {{\mathbf{c}}_l |{\mathbf{p}}_{il} } )$ is a
multinomial probability mass function with a parameter vector~${\mathbf{p}}_{il} $
and the number of trials equal to $c_{ \cdot l}$.

We attempt to minimize the dependence of the classifiers on the
underlying context in the database documents by basing the classifiers
on the conditional distributions of isocodes given letters and assuming
independence between the letters. By minimizing the contextual
dependence of the classifiers, we anticipate an increase in the
accuracy of our classifiers when applied to documents of unknown
writership with radically different context (when compared to the
modified ``London Letter'').

\subsection{Plug-In Naive Bayes Classifier}\label{sec4.1}
Given an estimate of ${\mathbf{p}}_{il} $, say, ${\mathbf{\hat p}}_{il}
$, we use the plug-in principle to estimate $P( {{\mathbf{c}}_l
|{\mathbf{p}}_{il} } )$ with $P( {{\mathbf{c}}_l |{\mathbf
{\hat p}}_{il} } )$ yielding the Plug-In Naive Bayes Classifier:
%
\begin{equation}\label{eq4.2}r( {{\mathbf{D}}_{uv} ,
{\mathbf{\hat P}}} ) = \Biggl\{ { \mathop{\arg\max}
 _{i \in\{ {1,2, \ldots, W} \}}
\prod _{l = 1}^L {P( {{\mathbf{n}}_{uvl} | {\mathbf{\hat
p}}_{il} } )} } \Biggr\},
\end{equation}
where ${\mathbf{\hat P}} = \{ {{\mathbf{\hat p}}_{il} \dvtx  i = 1,
2,  \ldots,W;  l = 1, 2, \ldots,L} \}$.
As suggested in \citet{r6}, we use a smoothed estimator of ${\mathbf
{p}}_{il} $,
%
\begin{equation}\label{eq4.3}\hat p_{iml} = \frac{{n_{i \cdot ml} + M^{ - 1} }}
{{n_{i \cdot \cdot l} + 1}}
\end{equation}
for $i = 1, 2,  \ldots,W; m = 1, 2, \ldots,M;$
and $l = 1, 2, \ldots,L.$ This estimate corresponds to the
expectation of the posterior distribution in the Dirichlet-Multinomial
Bayesian model, where the Dirichlet prior has $M$ shape parameters all
equal to~$M^{ - 1}$.

The classification procedure is as follows:
\begin{enumerate}[1.]
\item[1.] For each known writer in the database:
\begin{enumerate}[(a)]
\item[(a)] Estimate the conditional probability distribution of isocodes
using (\ref{eq4.3}).
\item[(b)] Use these conditional probability distributions to estimate the
likelihood, as in (\ref{eq4.1}), that an unknown document was written by a
given known writer.
\end{enumerate}

\item[2.]``Identify'' the unknown document as being written by the known
writer with the highest likelihood, as per (\ref{eq4.2}).
\end{enumerate}
Note that for a given writer in the database of writers, the Plug-In
Naive Bayes Classifier combines the individual documents associated
with the writer into one large writing sample.

This classifier is similar to the Naive Bayes Classifiers used in
authorship attribution by \citet{rv31} and \citet{rv32}. In \citet
{rv31}, the classifier is employed as a preliminary approach to a
fully Bayesian classification model. \citet{rv32} employ a
classifier similar to our Naive Bayes Classifier to study the potential
accuracy of different types of features in authorship attribution. In
authorship attribution applications, classes of words play a synonymous
role to that of letters in our work. The ``word within class'' plays a
role similar to that played by isocodes. \citet{rv31} noted that
their Naive Bayes Rule tends to possess extreme values of the posterior
log-odds of group membership. In the LOOCV performed in Section \ref{sec5}, a
similar behavior of the Plug-In Naive Bayes Classifier for writer
identification is observed.

\subsection{Chi-Squared Distance Classifier}\label{sec4.2}

In the handwriting biometric literature, a chi-squared style distance
metric for measuring the difference between two vectors of
probabilities has proven effective for nearest-neighbor style
classifiers. \citet{r8} compared Hamming, Euclid, Minkowski order 3,
Bhattacharya and chi-squared distance measure-based classifiers. The
chi-squared distance measure was found to outperform the other distance
measures. The nature of the handwriting data studied in \citet{r8} is
based on data-suggested categories that are determined by first
clustering bitmaps of either characters or parts of characters called
graphemes. A grapheme-based feature is classified into one of $k$
clusters, thus reducing an entire document into a single vector of
cluster proportions. Bulacu then uses a nearest-neighbor classifier to
predict the writer of a document of unknown writership. By working with
just proportions and not the counts, this type of classification scheme
effectively ignores the context and size of the document, which limits
the accuracy of the classifier when applied to small documents. The
Bulacu classifiers have been studied extensively and have been
demonstrated to be very effective in a broad range of applications
where the size of the documents is relatively large.

Based on Bulacu's research, we developed a version of the chi-squared
statistic that is applicable under the assumptions mentioned in the
introduction to this section. The basic approach is to apply a
chi-squared statistic to the vector of counts by letter and then
combine the chi-squared statistics across letters by taking advantage
of the independence assumption. However, before we can combine the
chi-squared statistics across letters, we will need to have a weighting
scheme that takes into account the relative information we have on each
letter. A natural way of doing this is to use the Pearson's chi-squared
test statistic.

To construct a score measuring the similarity between two documents
(i.e., a~similarity score), for each letter we calculate Pearson's
chi-squared statistic between the two vectors of isocode counts. This
results in a degrees of freedom and chi-squared statistic for each
letter used in both handwritten documents. The degrees of freedom and
the chi-squared statistics are summed across letters. As a heuristic,
the sum of chi-squared statistics is evaluated as a realization of a
chi-squared random variable with degrees of freedom equal to the sum of
degrees of freedom from the individual test statistics. If the
distributions are different, the resulting chi-squared statistic will
tend to be larger than when the distributions are the same. The
similarity score is the corresponding ``$p$-value'' to the omnibus
chi-squared statistic and degrees of freedom. This is repeated for each
known writer and the unknown document is associated with the writer
that has the largest \mbox{$p$-value}.

The classification procedure is as follows:

\begin{enumerate}[1.]
\item[1.] For each of the sample documents of known writership in the database:
\begin{enumerate}[(a)]
\item[(a)] Conditional on each letter, calculate Pearson's chi-squared
statistic on a two-way table of counts with two rows. The two rows
represent two documents: the sample document in the database and the
unknown document. The columns represent the various isocodes used to
write a given letter.
\item[(b)] Sum these chi-squared statistics across all letters.
Additionally, because the documents may use different numbers of
isocodes to represent different letters, sum the degrees of freedom
associated with the different chi-squared statistics.
\item[(c)] Using a chi-square distribution approximation with the summed
degrees of freedom, calculate an approximate $p$-value associated with
the summed statistic.
\end{enumerate}
\item[2.] ``Identify'' the unknown document as being written by the known
writer with the largest $p$-value.
\end{enumerate}

The Chi-Squared Distance Classifier is appropriate for
nearest-neighbor type applications where it may not be reasonable to
combine documents within a writer into a pooled writing sample.
Pearson's chi-squared statistics are commonly used in author
attribution to measure the discrepancy between the two sets of
frequencies of textual measurements associated with two documents. The
common approach is to exclude a text as having been written by a
specific author on the basis of an appropriate goodness-of-fit test
statistic. [For an example of this approach using Pearson's chi-squared
statistic, see \citet{rv33}.] However, chi-squared type statistics
have also been used as classifiers for author attribution studies. This
approach is to identify a text with an unknown author as having been
written by the author of the text with the smallest chi-squared
statistic. [See \citet{rv34} for an example.]

\subsection{Kullback--Leibler (KL) Distance Classifier}\label{sec4.3}
The final classifier is based on a symmetric version of the KL distance
[\citet{r7}]. The KL distance is a natural measure of the association
between two discrete distributions defined on the same sample space.
For two vectors of probabilities, ${\mathbf{q}}_1 $
and ${\mathbf{q}}_2 $ $ \in\mathbb{R}^M $, define the symmetric
KL-distance as
\[ \mathit{ KL}( {{\mathbf{q}}_1 ,{\mathbf{q}}_2 } ) =
{2}^{- 1} \sum _{m = {\mbox{1}}}^{M} {\biggl[ {q_{2m} {\operatorname{ln}}\frac{{q_{2m} }}{{q_{1m} }} + q_{1m} {\operatorname{ln}}\frac{{q_{1m} }}
{{q_{2m} }}} \biggr]}.
\]

The classification procedure is as follows:
\begin{enumerate}[1.]
\item[1.]
For the $j$th document from the $i$th writer in the database:
\begin{enumerate}[(a)]
\item[(a)] Estimate the conditional probability distribution of the isocodes
for the $l$th letter using ${\mathbf{\hat p}}_{il}^j = ( {\hat
p_{iml}^j } )_{M \times1}$, $l = 1, 2, \ldots,L$, where
$\hat p_{iml}^j $ is defined analogously to (\ref{eq4.3}).
\item[(b)] For each letter $l$, calculate the KL distance comparing the
conditional distribution for sample document $j$ from the $i$th
writer to the conditional distribution for the $u$th unknown document:
${\mathbf{\hat p}}_{ul} = ( {\hat p_{uml} } )_{M \times1}$,
$l = 1, 2, \ldots,L$, where $\hat p_{uml} $ is defined
analogously to (\ref{eq4.3}).
\item[(c)] Sum the distances across letters:
\[
\Delta( {u, i,  j} ) = \sum _{l = 1}^L {\mathit{ KL}(
{{\mathbf{\hat p}}_{il}^j ,{\mathbf{\hat p}}_{ul} } )}.
\]
\end{enumerate}
\item[2.] ``Identify'' the unknown document as being written by the
$i$th known writer if $\Delta( {u, i,  j} )$ is the
smallest value among $\{\Delta({u, i,  j})$, $i = 1,
2,\ldots, W, j=1, 2, \ldots, J_{i} \}$.
\end{enumerate}
As with the Chi-Squared Distance Classifier, the Kullback--Leibler
Distance Classifier is particularly appropriate for nearest-neighbor
type applications where it may not be reasonable to combine documents
within a writer into a pooled writing sample.

\section{Leave-one-out cross-validation}\label{sec5}

To evaluate these classifiers, a LOOCV scheme is implemented. For the
Plug-In Naive Bayes Classifier, each document in the database is
``left-out'' and the classifier $r( {  \cdot ,{\mathbf{\hat
P}}} )$ is constructed with the remaining documents. The left-out
document is then treated as a document of unknown writership and the
writership is predicted as $r( {\mathbf{D}_{uv} ,{\mathbf{\hat
P}}} )$. The single document from writer 16 was not used in
cross-validation. However, writer 16 was still a potential candidate
writer for other test documents. The accuracy of the classifier is
estimated by the number of times it correctly identifies the writership
of the left-out document. The Plug-In Naive Bayes Classifier correctly
identifies all documents.

A similar scheme is used to evaluate the Chi-Squared and
Kullback--Leibler Distance Classifiers. Each document in the data set
is ``left-out'' and treated as a document of unknown writership. Both
of these classifiers incorrectly classified the same single document,
which corresponds to estimated accuracy of $99.66\%$.

\section{Simulation}\label{sec6}
Based on the results of the LOOCV, our three classifiers are
effectively equal with close to $100\%$ accuracy when applied to the
full modified ``London Letter.'' To distinguish between the accuracy of
the three classifiers, we can stress the algorithms by giving them less
information. One of the properties that we would like our classifiers
to possess is high accuracy for unknown documents of relatively small size.

The natural way of exploring this would be to draw a subsample from the
set of observed characters in a given left-out writing sample. However,
due to the small size of some of the processed writing samples, the
possible document sizes that a subsampling approach could explore would
be limited. Additionally, a subsampling approach would give us
approximately the same proportion of letters in the documents in the
database and in the left-out document. It has been noted that having
the same context in both the unknown document and the database
documents affects the accuracy of the classifiers [\citet{r3}].

In the authorship attribution study of \citet{rv35}, a modified
LOOCV approach was proposed and implemented to estimate the accuracy of
their classifiers. This approach entails leaving out an entire body of
work from a single author and then classifying each of the blocks of
text within that body of work. We will implement a similar approach to
stress the ability of our classifiers to correctly assign writership of
a given writing sample. Due to the small writing sample size of some of
the handwritten documents, we are unable to look at individual blocks
of writing. In place of looking at the individual blocks of writing, a
parametric approach is used to simulate a random document from the
left-out document to be classified.\looseness=1

To generate a random document, predictive distributions are
constructed. A~Poisson distribution is used to determine the overall
frequency of occurrence of each letter observed in the left-out
document. A~multinomial distribution is used to determine the isocode
to be associated with an occurrence of a letter. All three of the
classifiers rely, in part, on an underlying assumption that for each
observed letter, the letter-dependent conditional distribution of
isocodes is multinomial. A~vector of proportions is estimated from the
left-out modified ``London Letter'' analogous to (\ref{eq4.3}). Then, for each
letter (say, the $l$th) observed in the left-out document, $x_l $
isocodes are sampled from the $l$th letter's predictive
distribution. We do not generate characters in the random document for
letters that are unobserved in the left-out document.

For the simulations presented in this paper, the means of the Poisson
random variables are $\mu = 1,  1.5$ and $2$. For each left-out
document, three random documents are generated at each mean value for a
total of nine random documents. For a single random document, the mean
value of the Poisson random variables is held constant across all
observed letters in the left-out document. The random generation of the
number of times we observe a given letter effectively generates a
document with radically different content than that of the original
modified ``London Letter.'' It should be noted that the nature of the
random document generation is forcing the isocode counts across letters
to be independent, which is one of the assumptions made in the
construction of the classifiers in Section \ref{sec4}.

\begin{figure}

\includegraphics{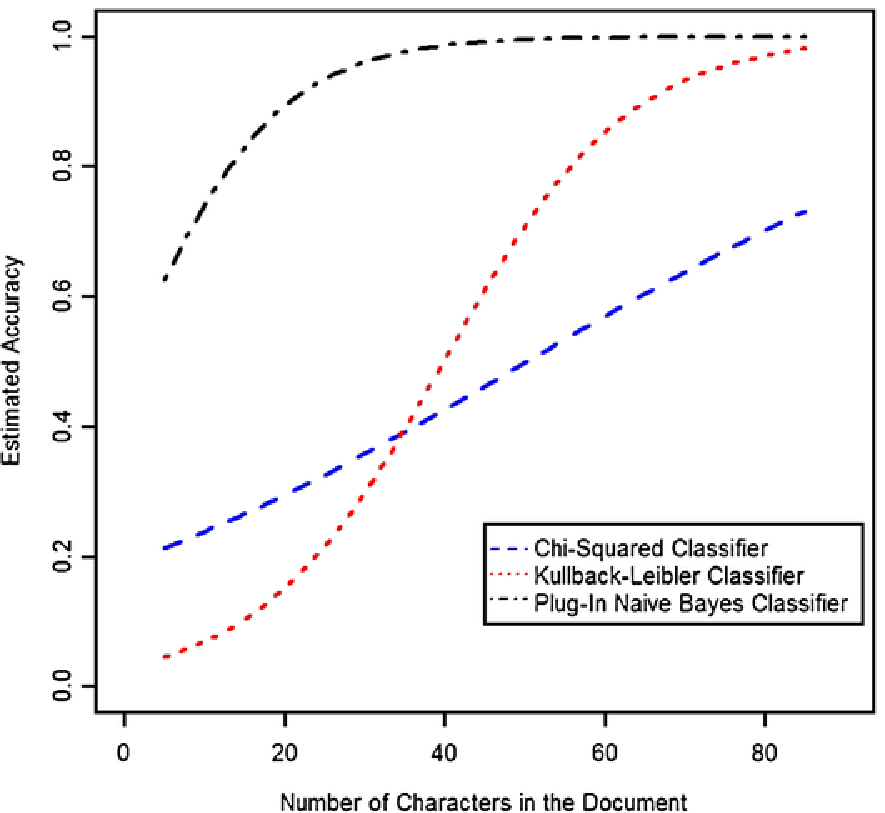}

\caption{The estimated accuracy of the classifiers as a function of the
number of characters in a document of unknown writership.}
\label{Figure4}
\end{figure}

Once a random document has been generated, a classifier predicts its
writership based on the other documents not used to generate it. To
summarize the results, a simple linear logistic regression is used to
predict the accuracy as a function of document size. The results are
summarized in Table \ref{parset} and Figure \ref{Figure4}.

\begin{table}
\caption{Summary of classifier accuracy. The first column, titled
number of characters, refers to the range in the number of characters
in the pseudo-documents. The number of pseudo-documents column refers
to the number of pseudo-documents of the size stated in the number of
characters column. The last three columns refer to the proportion of
pseudo-documents that are correctly identified by the given classifier:
`CS' for the Chi-Squared Distance Classifier, `KL' for the
Kullback--Leibler Distance Classifier, and `NB' for Plug-In Naive Bayes
Classifier}\label{parset}
\begin{tabular*}{\textwidth}{@{\extracolsep{\fill}}lcccc@{}}
\hline
\multicolumn{1}{@{}l}{\multirow{2}{45pt}[-6pt]{\textbf{Number of characters}}} &
\multicolumn{1}{c}{\multirow{2}{75pt}[-6pt]{\centering\textbf{Number of pseudo-documents}}} & \multicolumn{3}{c@{}}{\textbf{Accuracy}}
\\[-6pt]
 &  & \multicolumn{3}{c@{}}{\hrulefill} \\
&  & \multicolumn{1}{c}{\textbf{CS}} &
\multicolumn{1}{c}{\textbf{KL}} & \multicolumn{1}{c@{}}{\textbf{NB}} \\
\hline
{({0, 20}]} & {638} & 0.263 & {0.150} &
{0.840} \\
{({20, 30}]} & {829} & 0.328 & {0.217} &
{0.917} \\
{({30, 40}]} & {637} & 0.369 & {0.389} &
{0.980} \\
{({40, 50}]} & {347} & 0.441 & {0.637} &
{0.983} \\
{({50, 83}]} & {177} & 0.542 & {0.819} &
{1.000} \\
\hline
\end{tabular*}
\end{table}

Table \ref{parset} and Figure \ref{Figure4} suggest that the Plug-In
Naive Bayes Classifier has the highest accuracy of the three
classifiers. The Plug-In Naive Bayes Classifier achieves a $95\%$
accuracy rate for random documents of around 30 characters compared
with 70 characters for the Kullback--Leibler Distance Classifier (see
Figure~\ref{Figure4}). The performance of the Chi-Squared Distance
Classifier seems to suffer when applied to small documents.

The Dirichlet-Multinomial model has the effect of smoothing the
likelihood associated with each document. In the Kullback--Leibler
Distance Classifier, only a single document provides new information to
update the Dirichlet priors. This results in the Kullback--Leibler
Distance Classifier having the highest degree of smoothing [see (\ref{eq4.3})
and Section \ref{sec4.3}]. Due to pooling of the documents in the construction
of the Plug-In Naive Bayes Classifier, the effect of the Dirichlet
priors is washed out by the larger effective sample size. The
Chi-Squared Distance Classifier has no smoothing.

\section{Conclusions and future research}\label{sec7}
The proposed categorical classifiers have been demonstrated to have
near perfect accuracy, in terms of LOOCV error, when applied to the
``FBI 100'' data set. The random document simulations suggest that the
Plug-In Naive Bayes Classifier is the most efficient of the three
handwriting classifiers. It has a high identification accuracy rate for
documents of approximately 30 characters in size. The simulations
further suggest that the unknown document need not have the same text
as used for enrolling a writer into the database of writing samples for
the classifiers to have a high accuracy rate.

The accuracy of our classifiers applied to our current data set matches
or exceeds the accuracy rates of currently published handwriting
identification procedures, as summarized by \citet{r11}. The highest
level of accuracy of other researchers' classifiers requires larger
document sizes than the Plug-In Naive Bayes Classifier. However, to
compare the accuracy of our three classifiers with those proposed by
other researchers, all methods would need to be evaluated on a common
data set of documents.

A related problem to the writer identification problem addressed in
this paper concerns two competing hypotheses: ``the suspect wrote the
questioned document'' versus ``the suspect did not write the questioned
document.'' In this application, the evidence for deciding between the
two hypotheses is composed of both the handwriting samples collected
from the suspect (i.e., London Letters) and the document of unknown
writership. The classical approach of summarizing the value of the
evidence is to use a Bayesian likelihood ratio (also known as a Bayes
factor). [See the first three Chapters of \citet{rv36} for a
review.] If it is reasonable to assume that the distribution of
isocodes is independent across letters, then (\ref{eq4.1}) is an approximation
for the numerator of the Bayes factor (under the quantification
approach described in Section \ref{sec3}).

Alternatively, \citet{n4} provides a strategy to estimate the
likelihood ratio from an arbitrary biometric verification procedure.
Meuwly's approach is based on replacing the evidence (in the current
application, the writing exemplars collected from the suspect and
questioned document) with a score measuring the difference (or
similarity) between the suspect's exemplars and the questioned
document. The distribution of the score is then estimated under the two
competing hypotheses using appropriate databases of writing samples.
Both the Kullback--Leibler (KL) and the Chi-Squared Distance
Classifiers, proposed in Section \ref{sec4}, satisfy the necessary conditions of
a biometric verification procedure. The problem in handwriting is the
difficulty in creating a database of writing samples from the suspect
that is large enough to be able to accurately estimate the likelihood
of the observed score. We are currently exploring the potential of
applying resampling and subsampling approaches to a set of modified
``London Letters'' collected from the suspect to generate a
pseudo-database of writing samples. [See \citet{n2}.]

\section*{Authors' contributions}
CPS and LJD contributed equally and were involved in all stages of this
research and the development of the corresponding manuscript. ACL
contributed to the statistical coding for both LOOCV studies. DTG and
JJM suggested the research topic, provided the data (with support from
the Gannon Technologies Group) and contributed to the writing of the document.

\section*{Acknowledgments}
 JoAnn
Buscaglia provided the ``FBI 100'' data set and assisted in the
clarification of the original manuscript. The Gannon Technologies Group
gave computational support.
\vadjust{\eject}


\printaddresses

\end{document}